\documentclass[aps,prd,onecolumn,groupedaddress,showpacs,nofootinbib,amssymb]{revtex4-2}
\usepackage[dvips]{graphicx}
\usepackage{amssymb}
\usepackage{amsmath}
\usepackage{graphicx,,color}
\usepackage{amsfonts}
\usepackage{bm}
\usepackage{cancel}
\usepackage{comment}
\usepackage{floatflt}
\usepackage{slashed}
\newcommand\be{\begin{equation}}
\newcommand\ee{\end{equation}}

\allowdisplaybreaks[4]

\begin{document}

\title{Theoretical Probes of Higgs-Axion non-perturbative Couplings}
\author{V.K. Oikonomou,$^{1,2}$}\email{voikonomou@gapps.auth.gr;v.k.oikonomou1979@gmail.com}
\affiliation{$^{1)}$Department of Physics, Aristotle University of
Thessaloniki, Thessaloniki 54124, Greece \\ $^{2)}$L.N. Gumilyov
Eurasian National University - Astana, 010008, Kazakhstan}
%$^{2)}$ Laboratory for Theoretical Cosmology, International Center
%of Gravity and Cosmos, Tomsk State University of Control Systems
%and Radioelectronics  (TUSUR), 634050 Tomsk, Russia

 \tolerance=5000

\begin{abstract}
In this work we investigate the phenomenological implications of
several non-trivial axion-Higgs couplings, which cover most of the
possible non-perturbative scenarios. Specifically we consider the
combination of having higher order non-renormalizable Higgs-axion
couplings originating from a weakly coupled effective theory
combined with non-perturbative couplings of the form $\sim
\epsilon \Lambda_c^2|H|^2\cos (\frac{\phi}{f_a})$. Since we
consider the misalignment axion, the non-perturbative couplings
can be expanded in the form of a perturbation expansion in powers
of $\phi/f_a$, thus after the electroweak symmetry breaking, the
effective potential of the axion is drastically affected by these
terms. We investigate the phenomenological implications of these
terms for various values of the mass scale $\Lambda_c$, and some
scenarios are theoretically disfavored, while other scenarios with
non-perturbative Higgs-axion couplings of the form $\sim \epsilon
\Lambda_c^2|H|^2\cos (\frac{\phi}{f_a})$ with $\Lambda_c\sim
10^{-10}\times m_a$ and $m_a\sim 10^{-10}\,$eV, lead to a
characteristic pattern in the stochastic gravitational wave
background via the deformation of the background equation of state
parameter occurring at frequencies probed by the Einstein
Telescope. We also consider loop effects from the Higgs sector
caused by the term $\sim \epsilon \Lambda_c^2|H|^2\cos
(\frac{\phi}{f_a})$ if we close the Higgs in one loop.
\end{abstract}

%PACS numbers: 04.50.Kd, 95.36.+x, 98.80.-k, 98.80.Cq

\maketitle

\section{Introduction}

The next decade is anticipated by most theoretical physicists
since, many fundamental theories of theoretical physics will be
put into test. Specifically, inflation
\cite{inflation1,inflation2,inflation3,inflation4,inflation5}, one
of our cornerstone descriptions of the early Universe, will be
tested by the stage 4 Cosmic Microwave Background (CMB)
experiments \cite{CMB-S4:2016ple,SimonsObservatory:2019qwx}. The
future highly anticipated gravitational wave experiments will also
shed light on the primordial era of our Universe, by detecting
stochastic gravitational wave patterns
\cite{Hild:2010id,Baker:2019nia,Smith:2019wny,Crowder:2005nr,Smith:2016jqs,Seto:2001qf,Kawamura:2020pcg,Bull:2018lat,LISACosmologyWorkingGroup:2022jok}.
The chorus of stochastic gravitational wave detections has started
by the NANOGrav and other Pulsar Timing Array (PTA) collaborations
in 2023
\cite{nanograv,Antoniadis:2023ott,Reardon:2023gzh,Xu:2023wog}.
Now, regarding the stochastic gravitational wave background, the
NANOGrav 2023 signal cannot be explained by inflationary theories,
unless these have an abnormal large blue-tilted tensor spectral
index \cite{sunnynew}. There exist alternative possibilities, such
as low energy first order phase transitions with the transition
temperature being of the order $\sim \mathcal{O}(1)\,$GeV, since a
first order phase transition can generate bubble nucleation of the
new vacuum which can effectively produce stochastic gravitational
waves. This physical possibility applies to higher frequencies
too, where the electroweak phase transition can be probed by LISA,
the Einstein Telescope, the BBO and the DECIGO experiments, and a
significant amount of work has been produced examining the
possibility of having gravitational waves by first order phase
transitions during the radiation domination era
\cite{Apreda:2001us,Carrington:1991hz,Schabinger:2005ei,Kusenko:2006rh,McDonald:1993ex,Chala:2018ari,Davoudiasl:2004be,Baldes:2016rqn,Noble:2007kk,Zhou:2020ojf,
Weir:2017wfa,Hindmarsh:2020hop,Han:2020ekm,Child:2012qg,Fairbairn:2013uta,Caprini:2015zlo,Huber:2015znp,
Delaunay:2007wb,Barenboim:2012nh,Curtin:2014jma,Child:2012qg,Senaha:2020mop,Grojean:2006bp,Katz:2014bha}.
In most of the above scenarios, singlet extensions of the Standard
Model are considered, with the $SU(3)\times SU(2)\times U(1)$
singlet scalar being solely coupled with the Higgs sector
\cite{Profumo:2007wc,Damgaard:2013kva,Ashoorioon:2009nf,OConnell:2006rsp,Cline:2012hg,Gonderinger:2012rd,Profumo:2010kp,Gonderinger:2009jp,Barger:2008jx,
Cheung:2012nb,Alanne:2014bra,OConnell:2006rsp,Espinosa:2011ax,Espinosa:2007qk,Barger:2007im,Cline:2013gha,Burgess:2000yq,Kakizaki:2015wua,Cline:2012hg,
Enqvist:2014zqa,Barger:2007im}, or in some cases, Higgs self
couplings in the form of higher dimensional non-renormalizable
operators \cite{Chala:2018ari,Noble:2007kk,Katz:2014bha} can
achieve such a phase transition. Also in most of the cases, the
singlet extensions of the Standard Model serve as potential dark
matter candidate particles, since these scalars are perfect weakly
interacting massive particles (WIMPs). However, no WIMP has been
detected so far (2024), which casts some doubt on the very
existence of dark matter, without of course excluding the
possibility to find a WIMP in the next years, see
\cite{Arcadi:2024ukq} for a recent update on the WIMP scientific
status. Although formal relativistic materializations of Modified
Newtonian Dynamics are promising
\cite{Deffayet:2024ciu,Boran:2017rdn,Deffayet:2014lba,Deffayet:2011sk},
so far the dark matter mystery exists. The pessimistic scenario in
which dark matter is actually a particle that belongs to a truly
dark sector, completely uncoupled with the Standard Model sector,
is a true possibility. In that case, it is rather hard to verify
experimentally the existence of dark matter, at least in a direct
way. There is also however the scenario in which the Higgs sector
is weakly coupled to the dark matter particle. One strong dark
matter candidate, is the elusive axion particle
\cite{Preskill:1982cy,Abbott:1982af,Dine:1982ah,Marsh:2015xka,Sikivie:2006ni,Raffelt:2006cw,Linde:1991km,Co:2019jts,Co:2020dya,Barman:2021rdr,Marsh:2017yvc,Odintsov:2019mlf,Odintsov:2019evb,maxim,Anastassopoulos:2017ftl,Sikivie:2014lha,Sikivie:2010bq,Sikivie:2009qn,Masaki:2019ggg,Soda:2017sce,Soda:2017dsu,Aoki:2017ehb,Arvanitaki:2019rax,Arvanitaki:2016qwi,Machado:2019xuc,Tenkanen:2019xzn,Huang:2019rmc,Croon:2019iuh,Day:2019bbh,Oikonomou:2022ela,Oikonomou:2022tux,Odintsov:2020iui,Oikonomou:2020qah},
see also \cite{Semertzidis:2021rxs,Chadha-Day:2021szb} for reviews
and in addition an interesting simulation \cite{Buschmann:2021sdq}
for $\mu$eV range axions. The axion or some axion like particle,
is expected to have a significantly small mass, and several
experiments already exist targeting to detect the axion
\cite{BREAD:2021tpx}. Also proposals for the axion mass have
appeared in the literature \cite{Hoof:2022xbe,Li:2022pqa},
pointing to a light axion mass of the order $m_a\sim
\mathcal{O}(10^{-10})\,$eV, by using Gamma Ray Bursts
observational data. In this work we shall consider the
misalignment axion particle and we shall explore the
phenomenological implications of couplings of this light particle
with the Higgs sector. We shall be very cautious in choosing the
couplings, since renormalizable couplings may lead to a
thermalization of the axion, which is constrained by Higgs decays
in the Large Hadron Collider (LHC). The misalignment axion is a
non-thermal dark matter relic, so we will avoid studying
renormalizable couplings of the axion to the Higgs sector.
Specifically we shall consider the combination of dimension 4 and
dimension 6 higher order non-renormalizable couplings to the Higgs
particle and couplings of the form $\sim \epsilon
\Lambda_c^2|H|^2\cos (\frac{\phi}{f_a})$ where $\epsilon$ is some
constant smaller than unity, $\Lambda_c$ is an arbitrary scale,
$H$ is the Standard Model Higgs, $\phi$ is the axion and $f_a$ is
the axion decay constant. The effect of the simple case of having
only  dimension 4 and dimension 6 higher order non-renormalizable
couplings to the Higgs particle was considered in Ref.
\cite{Oikonomou:2023bah}, and in this work we extend the analysis
including couplings of the form $\sim \epsilon
\Lambda_c^2|H|^2\cos (\frac{\phi}{f_a})$, which were introduced in
\cite{Espinosa:2015eda}. Due to the misalignment mechanism, the
couplings  $\sim \epsilon \Lambda_c^2|H|^2\cos (\frac{\phi}{f_a})$
can be expanded in a perturbation expansion in terms of
$\frac{\phi}{f_a}$ and the effective potential of the axion is
altered after the electroweak phase transition. As we shall see,
the effects introduced by the couplings $\sim \epsilon
\Lambda_c^2|H|^2\cos (\frac{\phi}{f_a})$ are dramatic
phenomenologically, depending on the value of the mass scale
$\Lambda_c$. As we demonstrate, large mass scales of the order of
the Higgs mass are forbidden since they might lead to axion vacua
which are energetically equivalent to the Higgs vacuum. Also rich
phenomenology is provided for smaller values of the mass scale
$\Lambda_c$, and in some cases, the effects might generate a
detectable deformation in the stochastic gravitational wave
pattern during the radiation era.

This article is organized as follows: In section II we describe in
detail all the possible non-perturbative and higher order
non-renormalizable couplings of the Higgs to the axion sector. We
explain how the misalignment mechanism can provide the leading
order effect of the non-perturbative couplings in terms of a
perturbative expansion. After that we investigate the behavior of
the new axion effective potential at one loop order after the
electroweak symmetry breaking and we examine in detail the effect
of the non-perturbative couplings depending on the values of the
mass scale $\Lambda_c$. We also consider the thermalization
constraints and also constraints from the branching ratio of the
Higgs obtained by the LHC experiment. We also examine the
possibility to find observable effects on the stochastic
gravitational wave pattern for some values of the mass scale
$\Lambda_c$, generated by the deformation of the background
equation of state (EoS) during the radiation era, and occurs near
the frequency range probed by the Einstein Telescope. Also, the
quantum effects of couplings of the form $\sim \epsilon
\Lambda_c^2|H|^2\cos (\frac{\phi}{f_a})$, occurring by taking one
loop corrections of the Higgs are also examined in some detail.
Finally, the conclusions are presented in the end of the article.

\section{Non-perturbative Higgs-Axion Couplings, Higher Order non-renormalizable Operators and the Misalignment Mechanism}

Non-trivial couplings of the Higgs field with an axion-like
particle are quite frequently used in the literature in order to
generate a dynamical screening of the Higgs mass
\cite{Espinosa:2015eda}. These couplings are compatible with the
naturalness requirement, which is of fundamental importance in
theoretical physics. The theoretical aim of such terms is that
there is no need for new TeV physics in order to protect the Higgs
mass having large quantum corrections. There is a variety of
Higgs-axion terms that can be used in order to extract a specific
phenomenology, see for example \cite{Espinosa:2015eda}, however in
the present article we shall be interested in periodic couplings
of the form,
\begin{equation}\label{periodicterm}
\sim \epsilon \Lambda_c^2h^2\cos(\frac{\phi}{f_a})\, ,
\end{equation}
which in the context of Ref. \cite{Espinosa:2015eda} is used for
creating a potential barrier for the axion, where $\Lambda_c$ is
an arbitrary mass scale, and $\epsilon\leq 1$. Note that such
couplings are originating by the electroweak invariant terms of
the form $\sim |H|^2\cos (\frac{\phi}{f_a})$. Since we are not
interested in the phenomenology of Ref. \cite{Espinosa:2015eda},
we shall study the effects of such periodic terms given in Eq.
(\ref{periodicterm}) on the misalignment axion, taking also into
account higher order non-renormalizable couplings of the Higgs
particle to the axion. In all the cases, the thermalization issue
must be taken seriously into account, because the axion is a
non-thermal dark matter relic particle and also the branching
ratio of the Higgs must be very small in order to be compatible
with the LHC findings. We shall consider these issues in detail in
this section. We shall also provide the full effective potential
of the axion including the higher order non-renormalizable
couplings of the Higgs to the misalignment axion, and also
considering the one-loop corrected effective potential. Before
going into details of the model, let us discuss here the
misalignment axion mechanism which is very important and will play
a crucial role in the analysis.

Let us assume that the axion dynamical evolution is described by
the scenario known as misalignment axion
\cite{Marsh:2015xka,Co:2019jts}, in which there is a primordial
Peccei-Quinn $U(1)$ symmetry which is broken primordially, and
specifically in eras before the inflationary era. The axion field
emerges as the radial component of a primordial complex scalar
field which possesses the primordial $U(1)$ Peccei-Quinn symmetry.
Effectively, the axion during inflation is misaligned from the
minimum of its scalar potential,
\begin{equation}\label{axionpotentnialfull}
V_a(\phi )=m_a^2f_a^2\left(1-\cos \left(\frac{\phi}{f_a}\right)
\right)\, ,
\end{equation}
and its vacuum expectation value during inflation is quite large
of the order $\phi_i\sim f_a$, where $f_a>10^{9}\,$GeV and $f_a$
denotes the axion decay constant. The important feature of the
misalignment axion scenario is that during the inflationary era
and thereafter we have approximately $\phi/f_a<1$, therefore the
misalignment axion potential is approximated by the following,
\begin{equation}\label{axionpotential}
V_a(\phi )\simeq \frac{1}{2}m_a^2\phi^2\, .
\end{equation}
During the inflationary era the axion evolves towards the minimum
of its potential and when its mass $m_a$ becomes of the order of
the Hubble rate, then the axion eventually has reached the minimum
of its potential and it commences oscillations redshifting as cold
dark matter. Therefore in the post-inflationary evolution eras of
the Universe, the misalignment axion behaves as cold dark matter,
which is a non-thermal relic having no thermal contact with the
Standard Model particles. This also includes the Higgs particle,
so in our approach in which we want to respect the cold
non-thermal relic nature of the axion, the couplings of this
particle with the Higgs must be constrained so the
non-thermalization constraint is respected.

At this point let us present the full effective potential of the
axion at tree order, including all the possible couplings of it
with the Higgs particle, which includes dimension four and
dimension six higher order non-renormalizable couplings, and also
non-perturbative periodic couplings of the form given in Eq.
(\ref{periodicterm}). The full effective potential at tree order
is,
\begin{equation}\label{axioneightsixpotential}
V(\phi)=m_a^2f_a^2\left(1-\cos(\frac{\phi}{f_a})\right)-\lambda\frac{|H|^2\phi^4}{M^2}+g\frac{|H|^2\phi^6}{M^4}+
\Lambda_c^2|H|^2\cos(\frac{\phi}{f_a})\, .
\end{equation}
Since we are interested in the misalignment axion scenario, the
approximation $\phi\leq f_a$ is going to be applied at this point,
so the tree level potential at leading order in $\phi/f_a$ is,
\begin{equation}\label{axioneightsixpotentialmisalignment}
V(\phi)=\frac{1}{2}m_a^2\phi^2-\lambda\frac{|H|^2\phi^4}{M^2}+g\frac{|H|^2\phi^6}{M^4}+\frac{\epsilon\Lambda_c^2
|H|^2 \phi^6}{720 f_a^6}-\frac{\epsilon\Lambda_c^2 |H|^2
\phi^4}{24 f_a^4}+\frac{\epsilon\Lambda_c^2 |H|^2 \phi^2}{2
f_a^2}-\epsilon\Lambda_c^2 |H|^2\, .
\end{equation}
Note the presence of a $\phi$-independent term
$\sim-\epsilon\Lambda_c^2 |H|^2$ which serves as a dynamically
generated Higgs field mass term due to the misalignment axion
mechanism. We included this term in the potential just to show
that the misalignment approximation in the axion potential due to
the existence of the coupling (\ref{periodicterm}) leads to a
dynamically generated mass term for the Higgs field. We shall
further discuss this feature in a later section. Now, excluding
this Higgs-dependent term, the tree-level potential for the axion
field takes the form,
\begin{equation}\label{axioneightsixpotentialmisalignmentfinal}
V(\phi)=\frac{1}{2}m_a^2\phi^2-\lambda\frac{|H|^2\phi^4}{M^2}+g\frac{|H|^2\phi^6}{M^4}+\frac{\epsilon\Lambda_c^2
|H|^2 \phi^6}{720 f_a^6}-\frac{\epsilon\Lambda_c^2 |H|^2
\phi^4}{24 f_a^4}+\frac{\epsilon\Lambda_c^2 |H|^2 \phi^2}{2
f_a^2}\, .
\end{equation}
Now in the context of our approach, the Universe evolves and the
axion oscillates near the minimum of its scalar potential,
eventually however, when the temperature decreases to the order
$\sim \mathcal{O}(100)$GeV, the electroweak phase transition takes
place
\cite{Profumo:2007wc,Damgaard:2013kva,Ashoorioon:2009nf,OConnell:2006rsp,Cline:2012hg,Gonderinger:2012rd,Profumo:2010kp,Gonderinger:2009jp,Barger:2008jx,
Cheung:2012nb,Alanne:2014bra,OConnell:2006rsp,Espinosa:2011ax,Espinosa:2007qk,Barger:2007im,Cline:2013gha,Burgess:2000yq,Kakizaki:2015wua,Cline:2012hg,
Enqvist:2014zqa,Chala:2018ari,Noble:2007kk,Katz:2014bha} and the
Higgs field acquires a vacuum expectation value
$H^T=(\frac{\phi_2+i\phi_3}{\sqrt{2}},\frac{v+h+i\phi_1}{\sqrt{2}})$,
where $v=246\,$GeV is the electroweak vacuum. At this point, the
tree-level potential of the axion gets modified as follows,
\begin{equation}\label{axiontreelevelpotential}
V(\phi)=\frac{1}{2}m_a^2\phi^2-\lambda\frac{v^2\phi^4}{2
M^2}+g\frac{v^2\phi^6}{2 M^4}+\frac{\epsilon\Lambda_c^2 v^2
\phi^6}{1440 f_a^6}-\frac{\epsilon\Lambda_c^2 v^2 \phi^4}{48
f_a^4}+\frac{\epsilon\Lambda_c^2 v^2 \phi^2}{4 f_a^2}\, ,
\end{equation}
therefore the effective mass of the axion
$m_{eff}^2=\frac{\partial^2 V}{\partial \phi^2}$ reads,
\begin{equation}\label{effectiveaxionmass}
m_{eff}^2(\phi)=m_a^2+\frac{\Lambda_c^2 v^2 \phi^4}{48
f_a^6}-\frac{\Lambda_c^2 v^2 \phi^2}{4 f_a^4}+\frac{\Lambda_c^2
v^2}{2 f_a^2}+\frac{\phi^4 \left(15 g
v^2\right)}{M^4}-\frac{\phi^2 \left(6 \lambda v^2\right)}{M^2}\, .
\end{equation}
Now we can include the one-loop corrections to the effective
potential,
\begin{equation}\label{oneloopaxionzerotemperature}
V^{1}(\phi)=\frac{m_{eff}^4(\phi)}{64\pi^2}\left( \ln
\left(\frac{m_{eff}^2(\phi)}{\mu^2}\right)-\frac{3}{2}\right) \, ,
\end{equation}
and thus the total effective potential of the axion field up to
one-loop order is,
\begin{equation}\label{axiontreelevelpotential}
V^{1-loop}(\phi)=\frac{1}{2}m_a^2\phi^2-\lambda\frac{v^2\phi^4}{2
M^2}+g\frac{v^2\phi^6}{2 M^4}+\frac{\epsilon\Lambda_c^2 v^2
\phi^6}{1440 f_a^6}-\frac{\epsilon\Lambda_c^2 v^2 \phi^4}{48
f_a^4}+\frac{\epsilon\Lambda_c^2 v^2 \phi^2}{4
f_a^2}+\frac{m_{eff}^4(\phi)}{64\pi^2}\left( \ln
\left(\frac{m_{eff}^2(\phi)}{\mu^2}\right)-\frac{3}{2}\right)\, ,
\end{equation}
where $m_{eff}^2(\phi)$ is defined in Eq.
(\ref{effectiveaxionmass}). In the following we shall extensively
study the above potential for various interesting values of the
mass scale $\Lambda_c$. Now regarding the effective theory energy
scale $M$ appearing in the above equations, it characterizes the
energy scale at which this theory becomes active. We shall take
that scale to be $M=20\,$TeV, which is well above the current LHC
experiment operation energy which is nearly 13.5$\,$TeV
center-of-mass. The effective theory will be assumed to be weakly
coupled with the Wilson coefficients $\lambda$ and $g$ being of
the order $\lambda \sim \mathcal{O}(10^{-35})$ and $g\sim
\mathcal{O}(10^{-30})$.

Now, let us here extensively discuss an important issue having to
do with the peril of having the axion thermalized via interactions
of the form $\sim h^2\phi^2$ which exist in the effective
potential of the axion field  in Eq.
(\ref{axioneightsixpotentialmisalignmentfinal}). Note that the
Higgs tree-order potential is the following,
\begin{equation}\label{higgsaxiontreepotential}
V(H)=-m_H^2|H|^2+\lambda_H|H|^4\, ,
\end{equation}
and note that in the context of our work, in some cases, the term
$-m_H^2|H|^2$ can be dynamically generated by the misalignment
axion mechanism, however for completeness we include it here for
the shake of the thermalization argument. Note that $m_H=125$ GeV
is the Higgs boson mass, and also $\lambda_H$ denotes the Higgs
self-coupling, with
$\frac{v}{\sqrt{2}}=\left(\frac{-m_H^2}{\lambda_H}\right)^{\frac{1}{2}}$,
and also $v$ stands for the electroweak symmetry breaking scale
$v\simeq 246\,$GeV. Furthermore, $m_a$ denotes the axion mass,
which will be taken to be of the order $m_a\sim
\mathcal{O}(10^{-10})\,$eV. The weakly coupled effective
non-renormalizable operators do not affect the decays of the Higgs
particle to the axion sector, since the induced branching ratio is
small. Indeed experimentally what is expected is that the
branching ratio of the Higgs to the hidden scalar sector is
$\mathrm{BR}_{inv}<0.30-0.75$ at $95\%$CL \cite{Chung:2012vg},
therefore in our case, the non-renormalizable effective
interactions do not affect the branching ratio. However, the
interactions $\sim \lambda_I h^2\phi^2$ which exist in the
effective potential of the axion field  in Eq.
(\ref{axioneightsixpotentialmisalignmentfinal}) can affect the
branching ratio and indeed will if the parameter $\lambda_I$ does
not satisfy specific constraints. The couplings $\sim \lambda_I
h^2\phi^2$ are favored from phenomenology since the excess of
Higgs decay to diphotons can be enhanced by such couplings, thus
are plausible in the theory, however for the reasons we discussed,
the parameter $\lambda_I$ must be constrained. The decay rate of
the Higgs to the axion particle is \cite{Barger:2007im},
\begin{equation}\label{axionhiggsdecayrate}
\Gamma (h\to \phi \phi)=\frac{\lambda_I^2v^2}{32\pi
m_H}\sqrt{1-\frac{4m_a^2}{m_H^2}}\, .
\end{equation}
In our case, $\lambda_I$ is $\lambda_I=\frac{\Lambda_c^2}{2
f_a^2}$ which is significantly suppressed by the axion decay
constant. In any case the choices of $\Lambda_c$ must be such so
that the parameter $\lambda_I$ does not affect significantly the
branching ratio of the Higgs. There is another constraint
regarding the parameter $\lambda_I$ which must be taken into
account, having to do with the thermalization of the axion. The
axion is a cold dark matter relic and thus should not be
thermalized by the Higgs. However, an interaction of the form
$\lambda_I|h|^2\phi^2$ will thermalize the axion since, in a
ordinary radiation domination epoch, the thermalization rate is
approximated by $\Gamma_{th}\sim \lambda_I^2 T$ when the
temperature of the Universe is way above the Higgs mass, and when
the temperature becomes smaller than the Higgs mass, then the
thermalization rate becomes $\Gamma_{th}\sim \lambda_I^2
T^5m_H^{-4}$. Hence, the ratio of the thermalization rate over the
Hubble rate becomes maximized when we have approximately $T\sim
m_W$, with $m_W$ being the mass of the W boson. The thermalization
condition of dark matter scalar particles is $\lambda_I \geq
\sqrt{\frac{m_W}{M_p}}\sim 10^{-8}$ \cite{Burgess:2000yq}. Thus in
our case we must have $\lambda_I=\frac{\Lambda_c}{f_a}\ll 10^{-8}$
in order for the axion not to be thermalized via the interaction
$\sim \lambda_I h^2\phi^2$. In all the following cases which we
study in the next section, we shall take both the thermalization
constraint and the branching ratio constraint into account.

\subsection{Scenario I: Couplings of the Form $m_H^2|H|^2\cos (\frac{\phi}{f_a})$}

\begin{figure}
\centering
\includegraphics[width=20pc]{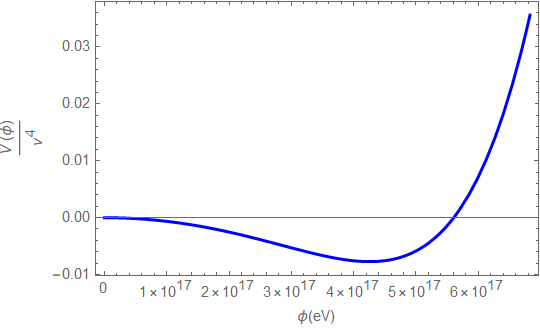}
\includegraphics[width=20pc]{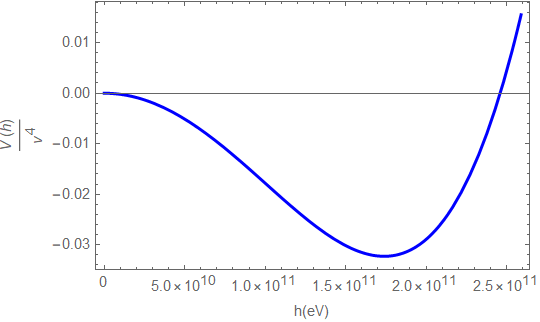}
\caption{The axion potential of Eq.
(\ref{axiontreelevelpotential}) for $\Lambda_c=m_H$ (left plot)
and the Higgs potential (right plot) after the electroweak
symmetry breaking.}\label{plot1}
\end{figure}

Let us consider a specific scenario for the value of $\epsilon
\Lambda_c$ appearing in the term  $\Lambda_c^2|H|^2\cos
(\frac{\phi}{f_a})$, and we assume that $\Lambda_c=m_H$. We refer
to this scenario as ``scenario I'' hereafter. Note that we shall
not take into account loop corrections induced from the Higgs
sector, thus we shall choose $\epsilon\sim \mathcal{O}(1)$ for
simplicity. Without taking into account quantum corrections from
the Higgs sector, let us also assume that the Higgs sector tree
level potential is given by,
\begin{equation}\label{higsspot1}
V(H)=\lambda_H|H|^4\, .
\end{equation}
Now, de to the misalignment mechanism of the axion, the term
$-m_H^2|H|^2\cos (\frac{\phi}{f_a})$ yields the Higgs field
dependent term $-m_H^2|H|^2$ at leading order, if we expand the
cosine function for $\phi\ll f_a$ which is exactly what the
misalignment mechanism describes. Thus in the context of scenario
I, the Higgs mass term is dynamically generated by Higgs-axion
couplings, and this term can be generated primordially, even
during the inflationary era, at which the axion is misaligned from
the minimum of its potential and slowly rolls to the minimum. This
is a rather interesting feature of the scenario I. Apart from
that, the axion dependent potential is given in Eq.
(\ref{axiontreelevelpotential}), thus in this section we will
focus on the phenomenological implications of such a potential
with the choice $\Lambda_c=m_H$. Firstly let us discuss the
behavior of the branching ratio and the thermalization issue for
$\Lambda_c=m_H$, in which case the coupling $\lambda_I$ entering
both the thermalization constraint and the branching ratio is
$\lambda_I=\frac{\Lambda_c^2}{2 f_a^2}$. In our case, we have
$\lambda_I\sim \mathcal{O}(10^{-15})$, so the thermalization
constraint $\lambda_I\leq 10^{-8}$ is satisfied and the same
applies for the decay rate which is of the order $\Gamma
(h\to\phi\phi)\sim \mathcal{O}(10^{-19})$eV. Now let us proceed to
the analysis of the axion effective potential, and in Fig.
\ref{plot1} we plot the effective potential for the axion given in
Eq. (\ref{axiontreelevelpotential}), which recall that is
generated after the electroweak symmetry takes place, and also we
quote for comparison the effective potential corresponding to the
Higgs potential.
As it can be seen, the axion potential is severely altered since a
new minimum is generated, but still the new vacuum respects the
misalignment mechanism which remains valid even with the dynamical
effects of the extra Higgs-axion couplings after the electroweak
symmetry breaking. More importantly, the Higgs vacuum is somewhat
deeper, but energetically the two vacua are almost equivalent.
Thus from a phenomenological point of view, scenario I is
theoretically disfavored because it leads to an additional vacuum
state in the Universe, which is equivalent to the Higgs vacuum. We
would then have two competing vacua in the Universe, which is
rather phenomenologically unacceptable and theoretically
unmotivated. Apart from that, the axion field as a dark matter
particle is basically eliminated since the new vacuum cannot decay
to the Higgs vacuum and thus the new mass of the axion is of the
order $\sim 0.47\,$GeV by using the values of the parameters with
$\Lambda_c=m_H$. Thus, although this change in its mass is not a
problem, the two competing vacua is an undesirable feature.

There is an additional reason which makes couplings of the form
$m_H^2|H|^2\cos (\frac{\phi}{f_a})$ undesirable both
phenomenologically and theoretically, apart from the reason we
just described. The reason is that if we include the loop
corrections effects at even one loop from the Higgs particle, the
axion acquires huge corrections from the Higgs sector. We shall
discuss this issue in the last section of this work. Let us
mention that the theoretically unappealing features of couplings
with $\Lambda_c\sim v$ were also discussed in Ref.
\cite{Espinosa:2015eda}.

\subsection{Scenario II: Couplings of the Form $m_{\nu}^2|H|^2\cos (\frac{\phi}{f_a})$ and $m_a^2|H|^2\cos (\frac{\phi}{f_a})$}

\begin{figure}
\centering
\includegraphics[width=20pc]{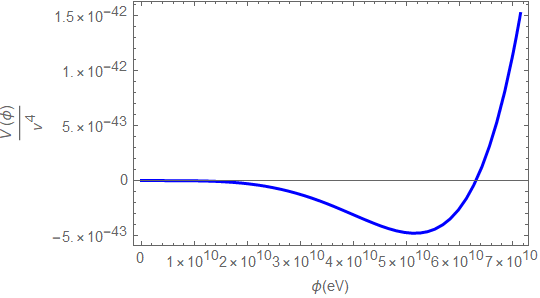}
\includegraphics[width=20pc]{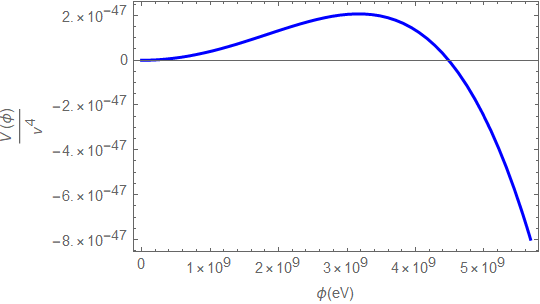}
\caption{The axion potential of Eq.
(\ref{axiontreelevelpotential}) for $\Lambda_c\sim
\mathcal{O}(m_{\nu})$.}\label{plot2}
\end{figure}

Now let us assume that the mass scale $\Lambda_c$ takes lower
values than the case studied in the previous section. We shall
consider two distinct scenarios, one in which case $\Lambda_c\sim
\mathcal{O}(m_{\nu})$ and one case in which $\Lambda_c\sim
\mathcal{O}(10^{-10}\times m_a)$, where $m_{\nu}$ is the neutrino
mass assumed to be of the order $m_{\nu}\sim 0.001\,$eV and the
axion mass is $m_a\sim 10^{-10}\,$eV. We shall refer to this case
as scenario II hereafter. Let us focus on the case $\Lambda_c\sim
\mathcal{O}(m_{\nu})$, in which case the coupling $\lambda_I$
entering both the thermalization constraint and the branching
ratio, $\lambda_I=\frac{\Lambda_c^2}{2 f_a^2}$ is of the order
$\lambda_I\sim \mathcal{O}(10^{-43})$, so the thermalization
constraint $\lambda_I\leq 10^{-8}$ is satisfied and also the decay
rate is of the order $\Gamma (h\to\phi\phi)\sim
\mathcal{O}(10^{-75})$eV. Now let us study the axion effective
potential in this case, so in Fig. \ref{plot2} we plot the axionic
effective potential given in Eq. (\ref{axiontreelevelpotential}),
with $\Lambda_c\sim \mathcal{O}(m_{\nu})$
In this case we have quite intriguing phenomenology because the
axion field develops a new energetically favorable minimum,
compared to that in the origin, and also a barrier exists between
the two vacua (see right plot). Thus the following
phenomenological picture is implied: The axion will vacuum
penetrate to the new vacuum via a first order phase transition
possibly, and will acquire instantly a new vacuum expectation
value. However, if we compare the axion potential with the Higgs
potential in Fig. \ref{plot1}, it is apparent that the axion
vacuum is not energetically favorable compared to the Higgs
vacuum, thus the axion vacuum will decay to the Higgs vacuum. Thus
the axion will return to the origin and the decay procedure will
be perpetually repeated. It is expected that the axion phase
transition can generate some imprint on the stochastic
gravitational wave background corresponding to an era after the
electroweak symmetry breaking, however we shall not go into
details for this scenario, which will be covered elsewhere.

\begin{figure}
\centering
\includegraphics[width=20pc]{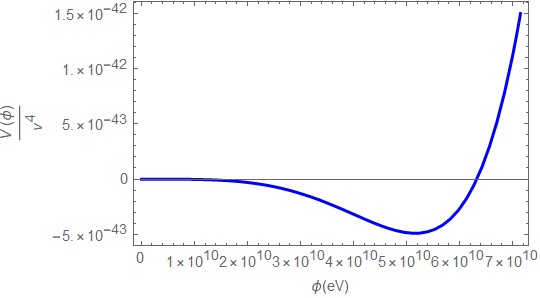}
\caption{The axion potential of Eq.
(\ref{axiontreelevelpotential}) for $\Lambda_c\sim
\mathcal{O}(10^{-10}\times m_a)$.}\label{plot3}
\end{figure}

\begin{figure}
\centering
\includegraphics[width=20pc]{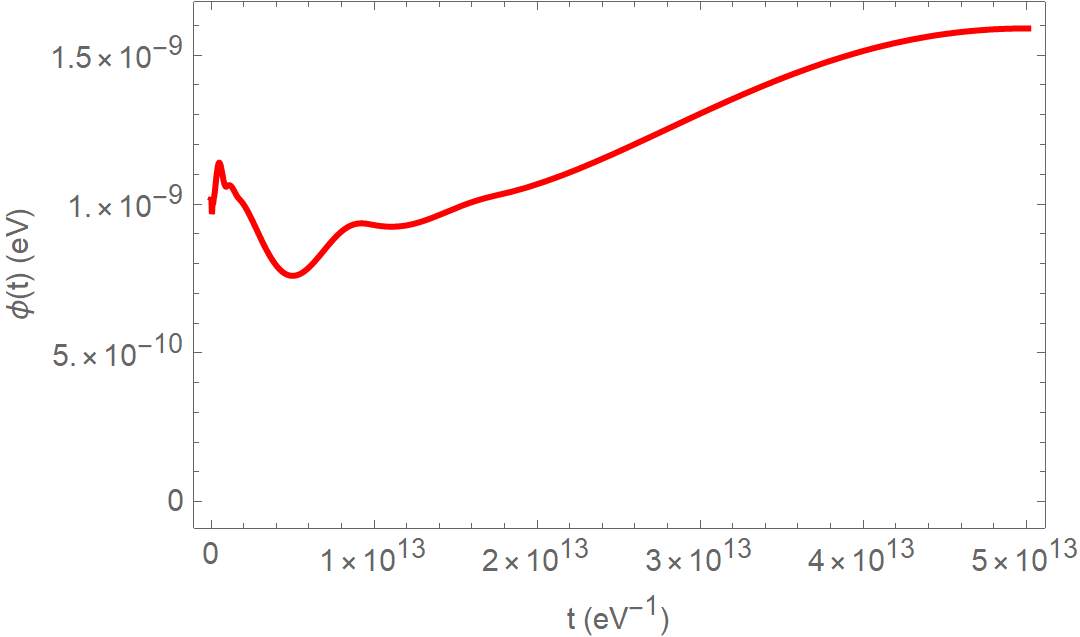}
\includegraphics[width=20pc]{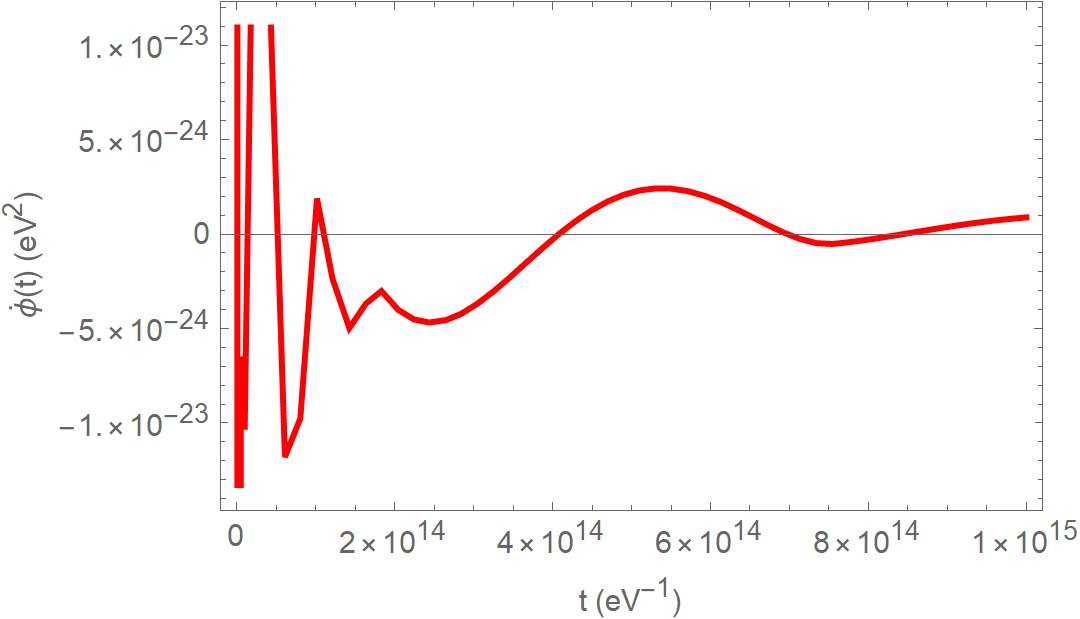}
\caption{The evolution of the axion $\phi(t)$ versus time (left
plot) at the origin of its effective potential, for the effective
potential given in Eq. (\ref{axiontreelevelpotential}) and the
evolution of $\dot{\phi}(t)$ versus time (right plot),
 for the initial conditions $\phi(0.00000001)=0.000000001$eV and
$\dot{\phi}(0.00000001)=0.01$eV$^2$.  This theory corresponds to
non-perturbative Higgs-axion couplings of the form $\sim \epsilon
\Lambda_c^2|H|^2\cos (\frac{\phi}{f_a})$ with $\Lambda_c\sim
10^{-10}\times m_a$ and $m_a\sim 10^{-10}\,$eV.}\label{plot4}
\end{figure}

\begin{figure}
\centering
\includegraphics[width=20pc]{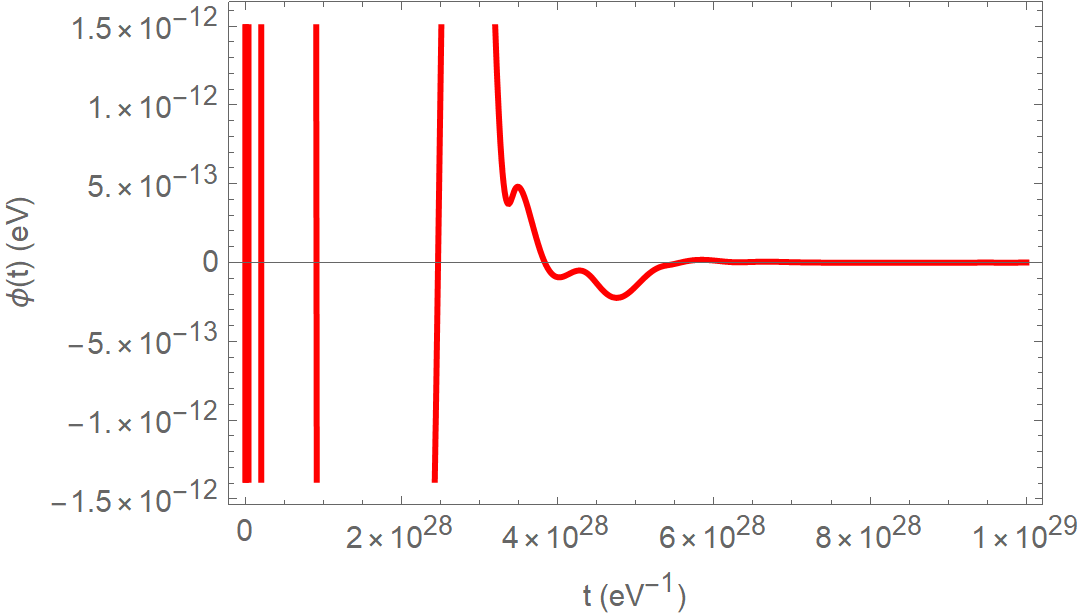}
\caption{The evolution of the axion $\phi(t)$ versus time at the
origin of its effective potential, for the effective potential
given in Eq. (\ref{axiontreelevelpotential}),
 for the initial conditions $\phi(0.00000001)=0$eV and
$\dot{\phi}(0.00000001)=0.01$eV$^2$. This behavior corresponds to
non-perturbative Higgs-axion couplings of the form $\sim \epsilon
\Lambda_c^2|H|^2\cos (\frac{\phi}{f_a})$ with $\Lambda_c\sim
10^{-10}\times m_a$ and $m_a\sim 10^{-10}\,$eV.}\label{plot5}
\end{figure}

Now let us turn our focus on the case that
$\Lambda_c=10^{-10}\times m_a$, with $m_a\sim 10^{-10}\,$eV. In
this case the coupling $\lambda_I$ entering both the
thermalization constraint and the decay rate of the Higgs to the
axion field due to interactions $\sim h^2\phi^2$ ratio,
$\lambda_I=\frac{\Lambda_c^2}{2 f_a^2}$ is of the order
$\lambda_I\sim \mathcal{O}(10^{-57})$, so in this case too the
thermalization constraint $\lambda_I\leq 10^{-8}$ is satisfied and
also the decay rate is of the order $\Gamma (h\to\phi\phi)\sim
\mathcal{O}(10^{-103})$eV. Now let us proceed to the study of the
axion effective potential in this case, so in Fig. \ref{plot3} we
plot the axionic effective potential given in Eq.
(\ref{axiontreelevelpotential}), with $\Lambda_c\sim
\mathcal{O}(10^{-10}\times m_a)$.
As we can see in this case, the axion develops a second minimum
which is energetically favorable compared to the minimum at the
origin, however no barrier exists between the two minima. In fact,
we observed that as the mass scale evolves from values
$\Lambda\sim m_{\nu}$ down to values $\Lambda_c\sim 10^{-15}\times
m_a$, the barrier exists but it is significantly small, and is
completely eliminated for values $\Lambda_c\sim 10^{-10}\times
m_a$ or lower. The phenomenology in this case is quite different,
because due to the existence of a second minimum with no barrier
separating the two minima, the axion oscillations at the origin
are interrupted. Thus after some considerable amount of time the
axion will cease to oscillate and will roll down the new potential
towards to the new minimum. Since the axion might be the sole
component of dark matter, this rolling of the axion down its
potential, will deform the background equation of state during
this rolling era, from $w=1/3$ which corresponds to a radiation
era, to a background EoS which is slightly smaller or larger than
$w=1/3$, depending on whether the axion slowly rolls or fast rolls
towards the new minimum. This background EoS deformation can have
some imprint on the energy spectrum of the stochastic
gravitational waves spectrum, as we evince shortly. However, when
the axion reaches the new minimum, the new vacuum will decay to
the Higgs vacuum, because the latter is energetically more
favorable compared to the axionic vacuum. Thus the axion returns
to the origin and the procedure is repeated perpetually. It is
hard to estimate how many times this procedure occurs, thus we
will assume that it occurs only once for simplicity. So for the
analysis of the gravitational wave energy spectrum, we shall focus
on the model with non-perturbative Higgs-axion couplings of the
form $\sim \epsilon \Lambda_c^2|H|^2\cos (\frac{\phi}{f_a})$ with
$\Lambda_c\sim 10^{-10}\times m_a$ and $m_a\sim 10^{-10}\,$eV.
Before going to that, let us try to investigate how the axion
oscillations are disturbed at the origin of its potential, with
the potential being given by Eq. (\ref{axiontreelevelpotential})
for $\Lambda_c\sim \mathcal{O}(10^{-10}\times m_a)$. Since the
deformation of the axion potential occurs after the electroweak
symmetry breaking, this means that the Hubble rate is described by
the radiation domination era value $H(t)=\frac{2}{t}$. The
evolution of the scalar field in terms of the cosmic time, is
described by the Klein-Gordon equation in a
Friedmann-Robertson-Walker background,
\begin{equation}\label{kleingordonfriedmann}
\ddot{\phi}+3H(t)\dot{\phi}+V'(\phi)=0\, ,
\end{equation}
so we shall numerically solve the above equation for the scalar
potential chosen as in Eq. (\ref{axiontreelevelpotential}) for
$\Lambda_c\sim \mathcal{O}(10^{-10}\times m_a)$ and for the rest
of the parameters chosen as in all the previous cases. We assume
that after the axion deformation of the potential, the axion
oscillates near the origin and has non-zero velocity, and
specifically that $\phi(0.00000001)=0.00001$eV and
$\dot{\phi}(0.00000001)=0.0001$eV$^2$. Thus, in Fig. \ref{plot4}
we present the evolution of $\phi(t)$ and of $\dot{\phi}(t)$ as a
function of the cosmic time.
As it can be seen in Fig. \ref{plot4}, the axion is destabilized
from its oscillations for $t\sim 10^{13}\,$eV$^{-1}=10^{-2}$sec,
which is quite fast, but it rolls with an incredibly small speed
towards the new minimum. This result however depends on the
initial conditions, so we shall assume that the evolution of the
axion towards the minimum of its potential can be done in two
ways, in a fast-roll way and in a slow-roll way and also that the
whole procedure occurs just once. This feature can be supported
numerically and it depends on the initial conditions of $\phi(t)$,
if for example $\phi(0)\sim 0$, then the axion oscillates near the
origin for quite long, even infinitely. This can be seen in Fig.
\ref{plot5} where it seems that the axion oscillates with a small
amplitude for nearly $t\sim 10^{15}$sec, which amounts for 0.3
billion of years. This result is robust regardless the value of
the speed of the axion, and it holds true even for relatively
large values of $\dot{\phi}$.
Thus, for models with non-perturbative Higgs-axion couplings of
the form $\sim \epsilon \Lambda_c^2|H|^2\cos (\frac{\phi}{f_a})$
with $\Lambda_c\sim 10^{-10}\times m_a$ and $m_a\sim
10^{-10}\,$eV, we assume that after the axion rolls its potential
once, and the new vacuum decays to the Higgs one, the axion
returns to the origin with its vacuum energy converted to kinetic
energy and it oscillates for a large amount of time before it will
start rolling again. Hence, the deformation of the background EoS
from its radiation domination value $w=1/3$ occurs only once. Now
let us investigate the effect of this background EoS deformation
on the energy spectrum of the stochastic gravitational waves. We
assume two distinct models for the inflationary era, one with the
standard red-tilted tensor spectral index, for example the $R^2$
model in the Jordan frame, or the corresponding model in the
Einstein frame, well-known as Starobinsky model, in which
$n_{\mathcal{T}}=-r/8$, and one model with a blue-tilted tensor
spectral index, such as one of the Einstein-Gauss-Bonnet models
developed in Refs. \cite{Oikonomou:2021kql}.
\begin{figure}[h!]
\centering
\includegraphics[width=40pc]{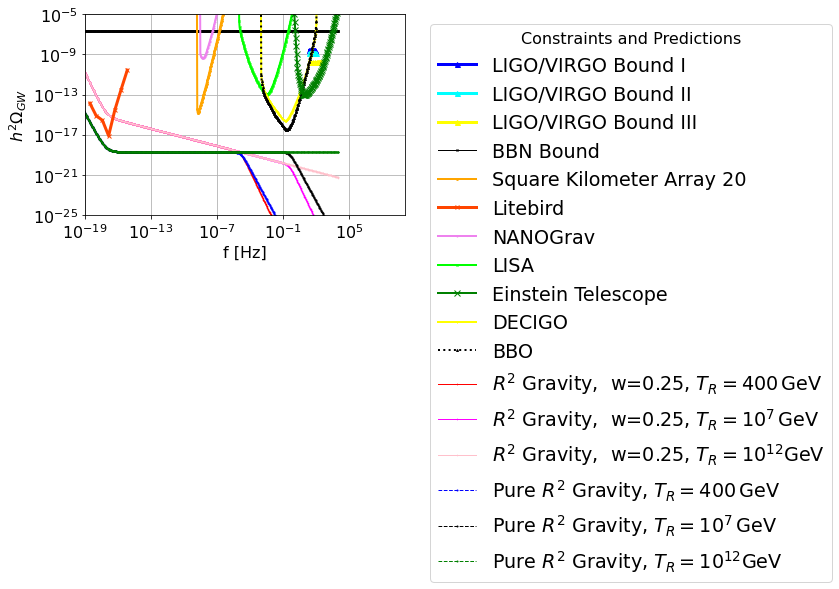}
\caption{The $h^2$-scaled gravitational wave energy spectrum for
an inflationary era described by $R^2$ theory with
$n_{\mathcal{T}}=-r/8$ and $r=0.003$, with the deformed background
EoS being and $w=0.25$ for frequencies probed by the Einstein
telescope $k_s=6.5\times 10^{16}$Mpc$^{-1}$, and for three
reheating temperatures $T_R=400\,$GeV, $T_R=10^7\,$GeV, and
$T_R=10^{12}\,$GeV. This phenomenology is obtained for
non-perturbative Higgs-axion couplings of the form $\sim \epsilon
\Lambda_c^2|H|^2\cos (\frac{\phi}{f_a})$ with $\Lambda_c\sim
10^{-10}\times m_a$ and $m_a\sim 10^{-10}\,$eV.}\label{plot6}
\end{figure}
For the analysis of the latter we will take that the tensor
spectral index takes the value $n_{\mathcal{T}}=0.37$ and also the
predicted tensor-to-scalar ratio is $r=0.003$, and the same
applies for the Starobinsky model, with $n_{\mathcal{T}}=-r/8$ in
this case however. Before proceeding, it is worth discussing in
brief the inflationary theories we shall consider and the
essential inflationary features of these. We start off with the
 $F(R)$ gravity theories which have the action,
\begin{equation}\label{action1dse}
\mathcal{S}=\frac{1}{2\kappa^2}\int \mathrm{d}^4x\sqrt{-g}F(R),
\end{equation}
and the corresponding field equations for a flat
Friedmann-Robertson-Walker metric are,
\begin{align}
\label{JGRG15} 0 =& -\frac{F(R)}{2} + 3\left(H^2 + \dot H\right)
F_R(R) - 18 \left( 4H^2 \dot H + H \ddot H\right) F_{RR}(R)\, ,\\
\label{Cr4b} 0 =& \frac{F(R)}{2} - \left(\dot H +
3H^2\right)F_R(R) + 6 \left( 8H^2 \dot H + 4 {\dot H}^2 + 6 H
\ddot H + \dddot H\right) F_{RR}(R) + 36\left( 4H\dot H + \ddot
H\right)^2 F_{RRR} \, ,
\end{align}
with $F_{RR}=\frac{\mathrm{d}^2F}{\mathrm{d}R^2}$, and also
$F_{RRR}=\frac{\mathrm{d}^3F}{\mathrm{d}R^3}$. Furthermore we
assume that the slow-roll conditions are valid,
\begin{equation}\label{slowrollconditionshubble}
\ddot{H}\ll H\dot{H},\,\,\, \frac{\dot{H}}{H^2}\ll 1\, .
\end{equation}
The inflationary indices, which capture the dynamics during the
inflationary era, namely $\epsilon_1$ ,$\epsilon_2$, $\epsilon_3$,
$\epsilon_4$, are \cite{Hwang:2005hb,reviews1,reviews6},
\begin{equation}
\label{restofparametersfr}\epsilon_1=-\frac{\dot{H}}{H^2}, \quad
\epsilon_2=0\, ,\quad \epsilon_3= \frac{\dot{F}_R}{2HF_R}\, ,\quad
\epsilon_4=\frac{\ddot{F}_R}{H\dot{F}_R}\,
 ,
\end{equation}
and using these, we can express the spectral index of the scalar
perturbations and also the tensor-to-scalar ratio in terms of
these \cite{reviews1,Hwang:2005hb},
\begin{equation}
\label{epsilonall} n_s=
1-\frac{4\epsilon_1-2\epsilon_3+2\epsilon_4}{1-\epsilon_1},\quad
r=48\frac{\epsilon_3^2}{(1+\epsilon_3)^2}\, .
\end{equation}
Furthermore, by using the Raychaudhuri equation, we have,
\begin{equation}\label{approx1}
\epsilon_1=-\epsilon_3(1-\epsilon_4)\, ,
\end{equation}
therefore in view of the slow-roll conditions, we get
approximately $\epsilon_1\simeq -\epsilon_3$, and in turn, the
scalar spectral index takes the following form,
\begin{equation}
\label{spectralfinal} n_s\simeq 1-6\epsilon_1-2\epsilon_4\, ,
\end{equation}
and also the tensor-to-scalar ratio becomes,
\begin{equation}
\label{tensorfinal} r\simeq 48\epsilon_1^2\, .
\end{equation}
Regarding the slow-roll index $\epsilon_4$, defined as
$\epsilon_4=\frac{\ddot{F}_R}{H\dot{F}_R}$ we get,
\begin{equation}\label{epsilon41}
\epsilon_4=\frac{\ddot{F}_R}{H\dot{F}_R}=\frac{\frac{d}{d
t}\left(F_{RR}\dot{R}\right)}{HF_{RR}\dot{R}}=\frac{F_{RRR}\dot{R}^2+F_{RR}\frac{d
(\dot{R})}{d t}}{HF_{RR}\dot{R}}\, ,
\end{equation}
and also by using,
\begin{equation}\label{rdot}
\dot{R}=24\dot{H}H+6\ddot{H}\simeq 24H\dot{H}=-24H^3\epsilon_1\, ,
\end{equation}
and in conjunction with Eq. (\ref{epsilon41}) we get,
\begin{equation}\label{epsilon4final}
\epsilon_4\simeq -\frac{24
F_{RRR}H^2}{F_{RR}}\epsilon_1-3\epsilon_1+\frac{\dot{\epsilon}_1}{H\epsilon_1}\,
.
\end{equation}
Also by using,
\begin{equation}\label{epsilon1newfiles}
\dot{\epsilon}_1=-\frac{\ddot{H}H^2-2\dot{H}^2H}{H^4}=-\frac{\ddot{H}}{H^2}+\frac{2\dot{H}^2}{H^3}\simeq
2H \epsilon_1^2\, ,
\end{equation}
the slow-roll index $\epsilon_4$ finally becomes,
\begin{equation}\label{finalapproxepsilon4}
\epsilon_4\simeq -\frac{24
F_{RRR}H^2}{F_{RR}}\epsilon_1-\epsilon_1\, .
\end{equation}
Accordingly the tensor spectral index can be evaluated using
\cite{Hwang:2005hb,reviews1,Odintsov:2020thl},
\begin{equation}\label{tensorspectralindexr2gravity}
n_{\mathcal{T}}\simeq -2 (\epsilon_1+\epsilon_3)\, ,
\end{equation}
hence by using  Eq. (\ref{finalapproxepsilon4}) we get,
\begin{equation}\label{tensorspectralindexr2ini}
n_{\mathcal{T}}\simeq -2 \frac{\epsilon_1^2}{1+\epsilon_1}\simeq
-2\epsilon_1^2\, .
\end{equation}
Regarding the $R^2$ model, we get,
\begin{equation}\label{r2modeltensorspectralindexfinal}
n_{\mathcal{T}}\simeq -\frac{1}{2N^2}\, ,
\end{equation}
therefore for $N=60$ we have $n_{\mathcal{T}}=-0.000138889$,
$n_s\simeq 0.963$ and $r\simeq 0.0033$, which we shall use for the
plots of the gravitational waves energy spectrum.

\begin{figure}[h!]
\centering
\includegraphics[width=40pc]{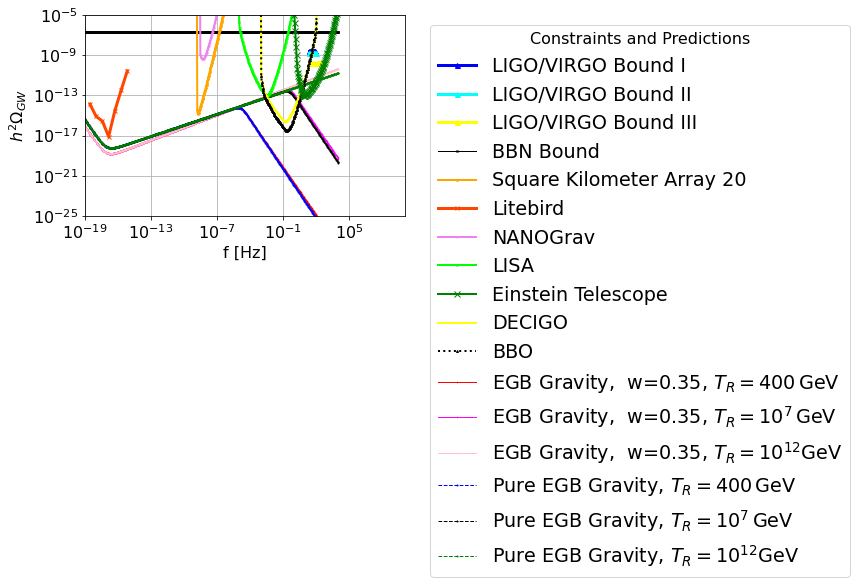}
\caption{The $h^2$-scaled gravitational wave energy spectrum for
an inflationary era described by an Einstein-Gauss-Bonnet theory
with $n_{\mathcal{T}}=0.37$ and $r=0.003$, with the deformed
background EoS being and $w=0.35$ for frequencies probed by the
Einstein telescope $k_s=6.5\times 10^{16}$Mpc$^{-1}$, and for
three reheating temperatures $T_R=400\,$GeV, $T_R=10^7\,$GeV, and
$T_R=10^{12}\,$GeV. This phenomenology for gravitational waves is
obtained for non-perturbative Higgs-axion couplings of the form
$\sim \epsilon \Lambda_c^2|H|^2\cos (\frac{\phi}{f_a})$ with
$\Lambda_c\sim 10^{-10}\times m_a$ and $m_a\sim
10^{-10}\,$eV.}\label{plot9}
\end{figure}

Regarding the Einstein-Gauss-Bonnet theories, the gravitational
action is,
\begin{equation}
\label{action} \centering
S=\int{d^4x\sqrt{-g}\left(\frac{R}{2\kappa^2}-\frac{1}{2}\partial_{\mu}\phi\partial^{\mu}\phi-V(\phi)-\frac{1}{2}\xi(\phi)\mathcal{G}\right)}\,
,
\end{equation}
where  $\mathcal{G}$ denotes the Gauss-Bonnet invariant which is
$\mathcal{G}=R^2-4R_{\alpha\beta}R^{\alpha\beta}+R_{\alpha\beta\gamma\delta}R^{\alpha\beta\gamma\delta}$
with $R_{\alpha\beta}$ and $R_{\alpha\beta\gamma\delta}$ are the
Ricci and Riemann tensors respectively. The propagation speed of
the gravitational tensor perturbations is required to be equal to
unity in Einstein-Gauss-Bonnet theories, hence by using this, the
slow-roll indices for the Einstein-Gauss-Bonnet theory become
\cite{Oikonomou:2021kql},
\begin{equation}
\label{index1} \centering \epsilon_1\simeq\frac{\kappa^2
}{2}\left(\frac{\xi'}{\xi''}\right)^2\, ,
\end{equation}
\begin{equation}
\label{index2} \centering
\epsilon_2\simeq1-\epsilon_1-\frac{\xi'\xi'''}{\xi''^2}\, ,
\end{equation}
\begin{equation}
\label{index3} \centering \epsilon_3=0\, ,
\end{equation}
\begin{equation}
\label{index4} \centering
\epsilon_4\simeq\frac{\xi'}{2\xi''}\frac{\mathcal{E}'}{\mathcal{E}}\,
,
\end{equation}
\begin{equation}
\label{index5} \centering
\epsilon_5\simeq-\frac{\epsilon_1}{\lambda}\, ,
\end{equation}
\begin{equation}
\label{index6} \centering \epsilon_6\simeq
\epsilon_5(1-\epsilon_1)\, ,
\end{equation}
with $\mathcal{E}=\mathcal{E}(\phi)$ and in addition
$\lambda=\lambda(\phi)$ are,
\begin{equation}\label{functionE}
\mathcal{E}(\phi)=\frac{1}{\kappa^2}\left(
1+72\frac{\epsilon_1^2}{\lambda^2} \right),\,\, \,
\lambda(\phi)=\frac{3}{4\xi''\kappa^2 V}\, .
\end{equation}
Thus, the inflationary observational indices can be obtained,
which are,
\begin{equation}
\label{spectralindex} \centering
n_{\mathcal{S}}=1-4\epsilon_1-2\epsilon_2-2\epsilon_4\, ,
\end{equation}
\begin{equation}\label{tensorspectralindex}
n_{\mathcal{T}}=-2\left( \epsilon_1+\epsilon_6 \right)\, ,
\end{equation}
\begin{equation}\label{tensortoscalar}
r=16\left|\left(\frac{\kappa^2Q_e}{4H}-\epsilon_1\right)\frac{2c_A^3}{2+\kappa^2Q_b}\right|\,
,
\end{equation}
where $c_A$ stands for the sound speed of the scalar
perturbations, which has the following form,
\begin{equation}
\label{sound} \centering c_A^2=1+\frac{Q_aQ_e}{3Q_a^2+
\dot\phi^2(\frac{2}{\kappa^2}+Q_b)}\, ,
\end{equation}
with,
\begin{align}\label{qis}
& Q_a=-4 \dot\xi H^2,\,\,\,Q_b=-8 \dot\xi H,\,\,\,
Q_t=F+\frac{Q_b}{2},\\
\notag &  Q_c=0,\,\,\,Q_e=-16 \dot{\xi}\dot{H}\, .
\end{align}
Thus, the tensor-to-scalar ratio and the tensor spectral index
become,
\begin{equation}\label{tensortoscalarratiofinal}
r\simeq 16\epsilon_1\, ,
\end{equation}
\begin{equation}\label{tensorspectralindexfinal}
n_{\mathcal{T}}\simeq -2\epsilon_1\left ( 1-\frac{1}{\lambda
}+\frac{\epsilon_1}{\lambda}\right)\, .
\end{equation}
A viable Einstein-Gauss-Bonnet model has the following
Gauss-Bonnet coupling function \cite{Oikonomou:2021kql},
\begin{equation}
\label{modelA} \xi(\phi)=\beta  \exp \left(\left(\frac{\phi
}{M}\right)^2\right)\, ,
\end{equation}
where $\beta$ is a dimensionless parameter and also $M$ has mass
dimensions $[m]^1$. The potential for this theory is derived to
be,
\begin{equation}
\label{potA} \centering V(\phi)=\frac{3}{3 \gamma  \kappa ^4+4
\beta  \kappa ^4 e^{\frac{\phi ^2}{M^2}}} \, ,
\end{equation}
where $\gamma$ a dimensionless an integration constant.
Accordingly, the slow-roll indices are,
\begin{equation}
\label{index1A} \centering \epsilon_1\simeq \frac{\kappa ^2 M^4
\phi ^2}{2 \left(M^2+2 \phi ^2\right)^2} \, ,
\end{equation}
\begin{equation}
\label{index2A} \centering \epsilon_2\simeq \frac{M^4
\left(2-\kappa ^2 \phi ^2\right)-4 M^2 \phi ^2}{2 \left(M^2+2 \phi
^2\right)^2}\, ,
\end{equation}
\begin{equation}
\label{index3A} \centering \epsilon_3=0\, ,
\end{equation}
\begin{equation}
\label{index5A} \centering \epsilon_5\simeq -\frac{4 \beta  \phi
^2 e^{\frac{\phi ^2}{M^2}}}{\left(M^2+2 \phi ^2\right) \left(3
\gamma +4 \beta  e^{\frac{\phi ^2}{M^2}}\right)} \, ,
\end{equation}
\begin{equation}
\label{index6A} \centering \epsilon_6\simeq -\frac{2 \beta  \phi
^2 e^{\frac{\phi ^2}{M^2}} \left(M^4 \left(2-\kappa ^2 \phi
^2\right)+8 M^2 \phi ^2+8 \phi ^4\right)}{\left(M^2+2 \phi
^2\right)^3 \left(3 \gamma +4 \beta  e^{\frac{\phi
^2}{M^2}}\right)} \, ,
\end{equation}
and in addition, the scalar spectral index, the tensor spectral
index and the corresponding tensor-to-scalar ratio are,
\begin{align}\label{spectralpowerlawmodel}
& n_{\mathcal{S}}\simeq -1-\frac{\kappa ^2 M^4 \phi
^2}{\left(M^2+2 \phi ^2\right)^2}+\frac{4 \phi ^2 \left(3 M^2+2
\phi ^2\right)}{\left(M^2+2 \phi ^2\right)^2}\\ & \notag
+\frac{4608 \beta ^2 \phi ^6 e^{\frac{2 \phi ^2}{M^2}} \left(6
\gamma  \phi ^2+16 \beta  e^{\frac{\phi ^2}{M^2}} \left(M^2+\phi
^2\right)+9 \gamma  M^2\right)}{\left(M^2+2 \phi ^2\right)^4
\left(3 \gamma +4 \beta  e^{\frac{\phi ^2}{M^2}}\right)^3} \, ,
\end{align}
and also
\begin{align}\label{tensorspectralindexpowerlawmodel}
& n_{\mathcal{T}}\simeq \frac{\phi ^2 \left(-4 \beta e^{\frac{\phi
^2}{M^2}} \left(M^4 \left(3 \kappa ^2 \phi ^2-2\right)+\kappa ^2
M^6-8 M^2 \phi ^2-8 \phi ^4\right)-3 \gamma \kappa ^2 M^4
\left(M^2+2 \phi ^2\right)\right)}{\left(M^2+2 \phi ^2\right)^3
\left(3 \gamma +4 \beta  e^{\frac{\phi ^2}{M^2}}\right)}
 \, ,
\end{align}
\begin{equation}\label{tensortoscalarfinalmodelpowerlaw}
r\simeq \frac{8 \kappa ^2 M^4 \phi ^2}{\left(M^2+2 \phi
^2\right)^2}\, .
\end{equation}
For the analysis and plots of the gravitational wave energy
spectrum we shall use the following values for the free parameters
$\mu=[22.0914,22.09147]$, $\beta=-1.5$, $\gamma=2$, for
approximately $N=60$ $e$-foldings, and for these we have
$n_{\mathcal{T}}=[0.378,0.379]$ and $r\sim 0.003$, which are what
we shall use for the gravitational wave analysis and plots. We
will also assume three distinct cases for the reheating
temperature, the following, $T_R=400\,$GeV, thus a low-reheating
temperature scenario, $T_R=10^7\,$GeV, thus an
intermediate-reheating temperature scenario, and finally a high
reheating temperature scenario with $T_R=10^{12}\,$GeV. We shall
also assume that the axion rolling towards the new minimum occurs
once during the radiation domination era, and at frequencies
probed by the Einstein telescope, thus for wavenumbers of the
order $k_{s}=6.5\times 10^{16}$Mpc$^{-1}$. Since there are two
ways for the axion to roll towards the new minimum of its
potential, one slow-roll and one fast-roll, the deformation of the
background EoS will be assumed to be either $w=0.25$, thus smaller
than $w=1/3$, or $w=0.35$, which is larger than the radiation
domination epoch value.
\begin{figure}[h!]
\centering
\includegraphics[width=40pc]{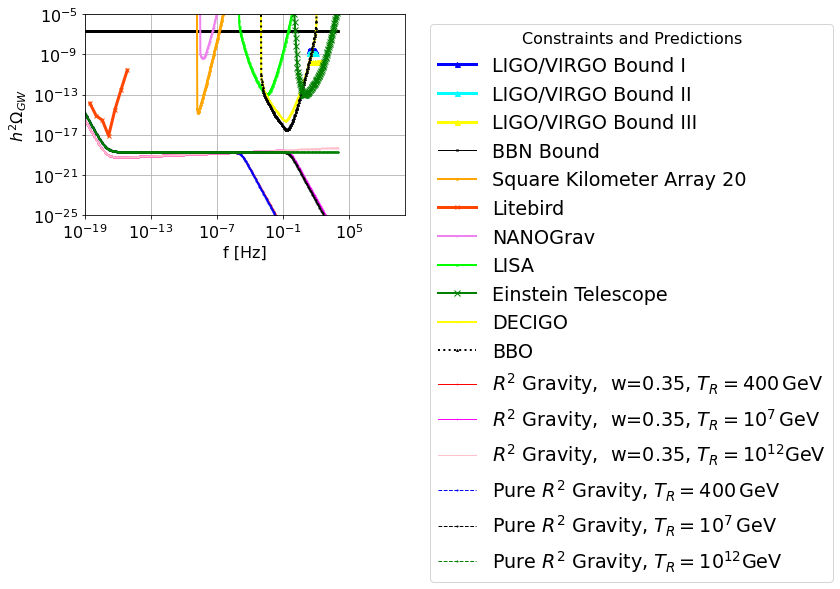}
\caption{The $h^2$-scaled gravitational wave energy spectrum for
an inflationary era described by $R^2$ theory with
$n_{\mathcal{T}}=-r/8$ and $r=0.003$, with the deformed background
EoS being and $w=0.35$ for frequencies probed by the Einstein
telescope $k_s=6.5\times 10^{16}$Mpc$^{-1}$, and for three
reheating temperatures $T_R=400\,$GeV, $T_R=10^7\,$GeV, and
$T_R=10^{12}\,$GeV. This phenomenological behavior is obtained for
non-perturbative Higgs-axion couplings of the form $\sim \epsilon
\Lambda_c^2|H|^2\cos (\frac{\phi}{f_a})$ with $\Lambda_c\sim
10^{-10}\times m_a$ and $m_a\sim 10^{-10}\,$eV.}\label{plot7}
\end{figure}
The effect of a background EoS deformation during the reheating
era on the energy spectrum of the gravitational waves is
quantified by a multiplication factor, $\sim
\left(\frac{k}{k_{s}}\right)^{r_c}$, where $r_c=-2\left(\frac{1-3
w}{1+3 w}\right)$ \cite{Gouttenoire:2021jhk}, where  $k_{s}$ is
the wavenumber at which the deformation of the background EoS
occurs.
\begin{figure}[h!]
\centering
\includegraphics[width=40pc]{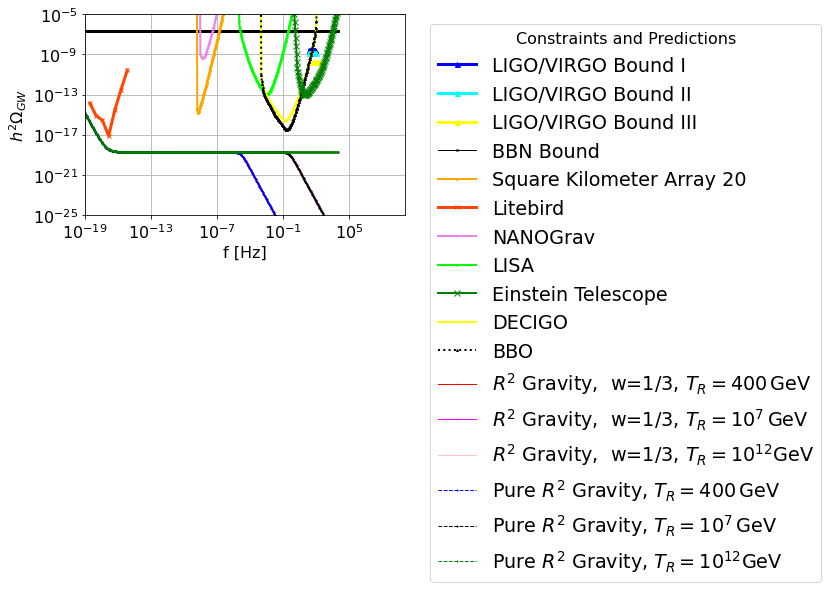}
\caption{The $h^2$-scaled gravitational wave energy spectrum for
an inflationary era described by $R^2$ theory with
$n_{\mathcal{T}}=-r/8$ and $r=0.003$, with the background EoS
being $w=1/3$ and for three reheating temperatures $T_R=400\,$GeV,
$T_R=10^7\,$GeV, and $T_R=10^{12}\,$GeV. This phenomenology is
obtained for models with non-perturbative Higgs-axion couplings of
the form $\sim \epsilon \Lambda_c^2|H|^2\cos (\frac{\phi}{f_a})$
with $\Lambda_c\sim 10^{-10}\times m_a$ and $m_a\sim
10^{-10}\,$eV.}\label{plot7a}
\end{figure}
This will multiply the $h^2$-scaled energy spectrum of the
primordial gravitational waves, thus including the effect of the
deformation of the background EoS,  the present day $h^2$-scaled
energy spectrum of the primordial gravitational waves takes the
form,
\begin{equation}\label{GWspecfRnewaxiondecay}
\Omega_{\rm gw}(f)=S_k(f)\times
\frac{k^2}{12H_0^2}r\mathcal{P}_{\zeta}(k_{ref})\left(\frac{k}{k_{ref}}
\right)^{n_{\mathcal{T}}} \left ( \frac{\Omega_m}{\Omega_\Lambda}
\right )^2
    \left ( \frac{g_*(T_{\rm in})}{g_{*0}} \right )
    \left ( \frac{g_{*s0}}{g_{*s}(T_{\rm in})} \right )^{4/3} \nonumber  \left (\overline{ \frac{3j_1(k\tau_0)}{k\tau_0} } \right )^2
    T_1^2\left ( x_{\rm eq} \right )
    T_2^2\left ( x_R \right )\, ,
\end{equation}
with $S_k(f)$,
\begin{equation}\label{multiplicationfactor1}
S_k(f)=\left(\frac{k}{k_{s}}\right)^{r_s}\, ,
\end{equation}
where $k_{ref}$ is the CMB pivot scale
$k_{ref}=0.002$$\,$Mpc$^{-1}$, and $n_{\mathcal{T}}$ and $r$ stand
for the tensor spectral index of the primordial tensor
perturbations and the tensor-to-scalar ratio. Also $T_{\rm in}$
denotes the temperature at the horizon reentry,
\begin{equation}
    T_{\rm in}\simeq 5.8\times 10^6~{\rm GeV}
    \left ( \frac{g_{*s}(T_{\rm in})}{106.75} \right )^{-1/6}
    \left ( \frac{k}{10^{14}~{\rm Mpc^{-1}}} \right )\, ,
\end{equation}
\begin{figure}[h!] \centering
\includegraphics[width=40pc]{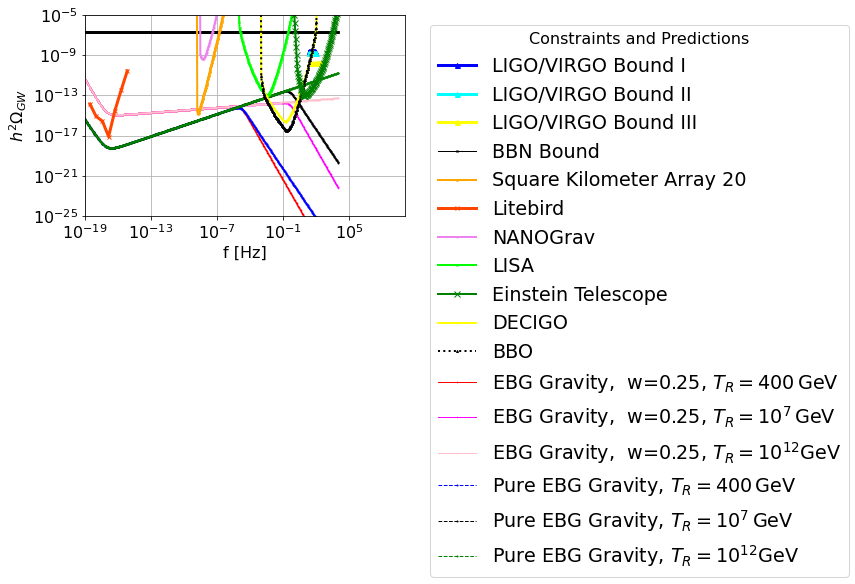}
\caption{The $h^2$-scaled gravitational wave energy spectrum for
an inflationary era described by an Einstein-Gauss-Bonnet theory
with $n_{\mathcal{T}}=0.37$ and $r=0.003$, with the deformed
background EoS being and $w=0.25$ for frequencies probed by the
Einstein telescope $k_s=6.5\times 10^{16}$Mpc$^{-1}$, and for
three reheating temperatures $T_R=400\,$GeV, $T_R=10^7\,$GeV, and
$T_R=10^{12}\,$GeV. This gravitational wave phenomenology is
obtained for non-perturbative Higgs-axion couplings of the form
$\sim \epsilon \Lambda_c^2|H|^2\cos (\frac{\phi}{f_a})$ with
$\Lambda_c\sim 10^{-10}\times m_a$ and $m_a\sim
10^{-10}\,$eV.}\label{plot8}
\end{figure}
and also the transfer function $T_1(x_{\rm eq})$ stands for,
\begin{equation}
    T_1^2(x_{\rm eq})=
    \left [1+1.57x_{\rm eq} + 3.42x_{\rm eq}^2 \right ], \label{T1}
\end{equation}
with $x_{\rm eq}=k/k_{\rm eq}$ and $k_{\rm eq}\equiv a(t_{\rm
eq})H(t_{\rm eq}) = 7.1\times 10^{-2} \Omega_m h^2$ Mpc$^{-1}$,
while the transfer function $T_2(x_R)$ is equal to,
\begin{equation}\label{transfer2}
 T_2^2\left ( x_R \right )=\left(1-0.22x^{1.5}+0.65x^2
 \right)^{-1}\, ,
\end{equation}
with $x_R=\frac{k}{k_R}$, and the wavenumber at reheating
temperature is,
\begin{equation}
    k_R\simeq 1.7\times 10^{13}~{\rm Mpc^{-1}}
    \left ( \frac{g_{*s}(T_R)}{106.75} \right )^{1/6}
    \left ( \frac{T_R}{10^6~{\rm GeV}} \right )\, ,  \label{k_R}
\end{equation}
where $T_R$ denotes the reheating temperature. Finally,
$g_*(T_{\mathrm{in}}(k))$ stands for \cite{Kuroyanagi:2014nba},
\begin{align}\label{gstartin}
& g_*(T_{\mathrm{in}}(k))=g_{*0}\left(\frac{A+\tanh \left[-2.5
\log_{10}\left(\frac{k/2\pi}{2.5\times 10^{-12}\mathrm{Hz}}
\right) \right]}{A+1} \right) \left(\frac{B+\tanh \left[-2
\log_{10}\left(\frac{k/2\pi}{6\times 10^{-19}\mathrm{Hz}} \right)
\right]}{B+1} \right)\, ,
\end{align}
where $A$ and $B$ are equal to,
\begin{equation}\label{alphacap}
A=\frac{-1-10.75/g_{*0}}{-1+10.75g_{*0}}\, ,
\end{equation}
\begin{equation}\label{betacap}
B=\frac{-1-g_{max}/10.75}{-1+g_{max}/10.75}\, ,
\end{equation}
with $g_{max}=106.75$ and $g_{*0}=3.36$. Furthermore
$g_{*0}(T_{\mathrm{in}}(k))$ can be calculated by using Eqs.
(\ref{gstartin}), (\ref{alphacap}) and (\ref{betacap}), by
replacing $g_{*0}=3.36$ with $g_{*s}=3.91$. Now let us proceed to
the analysis of the $h^2$-scaled energy spectrum of the primordial
gravitational waves including the effects of the deformation of
the background EoS. In Figs. \ref{plot6} and \ref{plot7} we plot
the $h^2$-scaled gravitational wave energy spectrum for an
inflationary era described by the Starobinsky model with
$n_{\mathcal{T}}=-r/8$ and $r=0.003$, with a deformed background
EoS $w=0.25$ (Fig. \ref{plot6}) and $w=0.35$ (Fig. \ref{plot7})
occurring for frequencies probed by the Einstein telescope
$k_s=6.5\times 10^{16}$Mpc$^{-1}$, for three reheating
temperatures $T_R=400\,$GeV, $T_R=10^7\,$GeV, and
$T_R=10^{12}\,$GeV, for both the pure and broken power-law cases.
In addition, in Fig. \ref{plot7a} we plot the energy spectrum of
the primordial gravitational waves for $w=1/3$ for ordinary $R^2$
gravity without the broken power-law for frequencies above the
Einstein Telescope, and with the broken power-law modifications.
Also in Figs. \ref{plot8} and \ref{plot9} we plot the $h^2$-scaled
gravitational wave energy spectrum for an inflationary era
described by an Einstein-Gauss-Bonnet theory with
$n_{\mathcal{T}}=0.37$ and $r=0.003$, with the deformed background
EoS being $w=0.25$ (Fig. \ref{plot8}) and $w=0.35$ (Fig.
\ref{plot9}) occurring in this case too for frequencies probed by
the Einstein telescope $k_s=6.5\times 10^{16}$Mpc$^{-1}$, and
again for three reheating temperatures $T_R=400\,$GeV,
$T_R=10^7\,$GeV, and $T_R=10^{12}\,$GeV, including the pure
Einstein-Gauss-Bonnet case without the broken power-law scenarios.
In all the plots we included the sensitivity curves from the most
prominent future gravitational wave experiments, and also the
constraints from the Big Bang Nucleosynthesis and the constraints
from LIGO-Virgo.
Let us analyze the resulting picture at this point and we consider
the $R^2$ case first. As it can be seen in Figs. \ref{plot7} and
\ref{plot7a}, the energy spectrum remains undetectable for
ordinary $R^2$ gravity without the broken power-law above the
Einstein Telescope frequencies, however for a broken power-law
above the Einstein Telescope frequencies with a deformed
background EoS equal to $w=0.25$, the energy spectrum of the
primordial gravitational waves is detectable by the Litebird
experiment as can be seen in Fig. \ref{plot6}. However, for the
case of $R^2$ gravity with a broken power-law above the Einstein
Telescope frequencies, with a deformed background EoS equal to
$w=0.35$, the energy spectrum of the primordial gravitational
waves is undetectable by all the future experiments and in fact
resembles the pure $R^2$ scenario. Thus for $w<1/3$ the
differences between the pure and deformed $R^2$ scenarios are more
pronounced. However for $w=1/3$ the broken power-law $R^2$ results
and the pure $R^2$ ones are indistinguishable, as it can be seen
in Fig. \ref{plot7a}. Regarding the Einstein-Gauss-Bonnet theories
with $n_{\mathcal{T}}=0.37$ and $r=0.003$, the signal for a broken
power-law above the Einstein Telescope frequencies, with a
deformed background EoS equal to $w=0.25$, the signal is
detectable only by the Litebird, and the SKA for a small reheating
temperature, by Litebird, LISA, BBO and DECIGO for intermediate
reheating temperatures, and by Litebird, LISA, BBO, DECIGO and the
Einstein Telescope for a large reheating temperature, as it can be
seen in Fig. \ref{plot8}. Finally, for Einstein-Gauss-Bonnet
theories with $n_{\mathcal{T}}=0.37$ and $r=0.003$, the signal for
a broken power-law above the Einstein Telescope frequencies, with
a deformed background EoS equal to $w=0.35$, the signal is
undetectable for a small reheating temperature, is detectable by
BBO and DECIGO for intermediate reheating temperatures, and
detectable by BBO, DECIGO and the Einstein Telescope for a large
reheating temperature, as it can be seen in Fig. \ref{plot9}.
Also, as in the $R^2$ case, the differences between pure
Einstein-Gauss-Bonnet theory and broken power-law
Einstein-Gauss-Bonnet theory is more pronounced when $w<1/3$. Now
the resulting picture is quite illuminating since it offers a
unique pattern for future stochastic gravitational wave detection.
The important information offered is the pattern of detection by
Litebird and the rest of the future experiments. The predictions
provide information about the tensor spectral index, if it is
positive or negative, and also provides information about the
reheating temperature, if it is high, intermediate or small.
Furthermore what is mentionable is the detection from Litebird in
both the red-tilted and blue-tilted inflation, especially for the
former. This information can be valuable for future stochastic
gravitational wave detections, since our analysis provides
specific patterns of detections. Thus a synergy between all the
future gravitational waves experiments can yield important
insights towards finding the correct underlying theory that
produces the future detected pattern of stochastic gravitational
waves.

\section{Corrections to the Misalignment Axion Potential due
to Higgs Loops}

Now let us consider another important feature brought in the
theory of axion-Higgs interactions by couplings of the form $\sim
\epsilon \L_c^2|H|^2\cos (\frac{\phi}{f_a})$. Now with these
couplings present, if one takes into account one-loop Higgs
corrections, the following quantum correction terms arise in the
theory \cite{Espinosa:2015eda},
\begin{equation}\label{quantumcorrection}
\epsilon \Lambda_c^4\cos (\frac{\phi}{f_a})\, ,
\end{equation}
at leading order in the dimensionless parameter $\epsilon$ which
is considered to be smaller than unity. So effectively if one
takes into account the misalignment approximation in the regime
$\phi\ll f_a$, the above quantum corrections result in the
following terms that must be added to the effective potential of
the axion, at leading order in $\phi$,
\begin{equation}\label{extraquantumterms}
V^{quantum}(\phi)=\epsilon \Lambda_c^4-\frac{\epsilon
\Lambda_c^4}{2f_a^2}\phi^2+\frac{\epsilon \Lambda_c^4}{24
f_a^4}\phi^4\, ,
\end{equation}
so the effective potential of the axion becomes,
\begin{align}\label{effectiveaxionmass}
& V(\phi)=\frac{\Lambda_c^2 v^2 \phi^4 \epsilon }{48
f_a^4}-\frac{\Lambda_c^2 v^2 \phi^2 \epsilon }{4 f_a^2}+m_a^2
\left(1-\cos \left(\frac{x}{f_a}\right)\right)+\frac{g v^2
\phi^6}{2 M^4}-\frac{\lambda  v^2 \phi^4}{2 M^2}\\ \notag &
-\frac{\Lambda_c^2 v^2 \phi^6 \epsilon }{1440
f_a^6}+\frac{\Lambda_c^4 \phi^4 \epsilon }{24
f_a^4}-\frac{\Lambda_c^4 \phi^2 \epsilon }{2
f_a^2}+\frac{m_{eff}^4 \left(\log \left(\frac{m_{eff}^2}{\mu
^2}\right)-\frac{3}{2}\right)}{64 \pi ^2}+\Lambda_c^4 \epsilon\, ,
\end{align}
where $m_{eff}^2$ in this case reads,
\begin{equation}\label{meffquantumcorr}
m_{eff}^2=\frac{\Lambda_c^2 v^2 \phi^4 \epsilon }{48
f_a^6}-\frac{\Lambda_c^2 v^2 \phi^2 \epsilon }{4
f_a^4}+\frac{\Lambda_c^4 \phi^2 \epsilon }{2
f_a^4}+\frac{\Lambda_c^2 v^2 \epsilon }{2 f_a^2}-\frac{\Lambda_c^4
\epsilon }{f_a^2}+\frac{\phi^4 \left(15 g
v^2\right)}{M^4}-\frac{\phi^2 \left(6 \lambda
v^2\right)}{M^2}+m_a^2\, .
\end{equation}
A noticeable feature brought by the quantum corrections is the
presence of a constant vacuum energy term $\sim \epsilon
\Lambda_c^4$. Now regarding the phenomenology, we shall assume
that $\epsilon\sim \mathcal{O}(0.1)$, so we consider all the
scenarios we discussed in the previous sections. The resulting
phenomenological picture is quite similar to the case where the
quantum corrections were not included. Particularly, for an
effective theory of $M\sim 20\,$TeV, the case with $\Lambda_c\sim
125\,$GeV leads to a new vacuum state for the axion which is
energetically equivalent with the Higgs vacuum, so this scenario
is rather phenomenologically unviable, plus a large vacuum energy
of the order $\sim \epsilon \Lambda_c^4$ is induced. Now the case
with $\Lambda_c\sim 10^{-3}\,$eV is quite interesting, since a
small constant vacuum energy term is induced and also the
phenomenological picture is the same with the case that the
quantum corrections were not included, except for the fact that
the barrier between the new and old axion vacua does not exist.
Thus the axion is free to roll down to its new minimum, a scenario
which was described in detail in the previous sections. The same
applies for the case $\Lambda_c\sim 10^{-20}\,$eV, in which case a
small  constant vacuum energy term is induced too.

\section*{Concluding Remarks and Discussion}

In this work we considered some of the most possible Higgs-axion
couplings and we investigated the phenomenological implications of
such couplings. Specifically, we assumed that the axion is coupled
to a weakly coupled effective theory with the energy scale $M$ of
the effective theory being of the order of $M\sim
\mathcal{O}(20)\,$TeV, and the Wilson  coefficients of the order
$\mathcal{O}(10^{-35})\,$TeV, in the presence of non-perturbative
couplings of the form $\sim \epsilon \Lambda_c^2|H|^2\cos
(\frac{\phi}{f_a})$ with $\Lambda_c$ being an arbitrary mass
scale. Due to the fact that we consider the misalignment axion,
during and in the post-inflationary era, the axion satisfies
$\phi\ll f_a$, thus the non-perturbative coupling can be expanded
in perturbation series in terms of $\phi/f_a$ and eventually we
may have a leading order approximation for the axion effective
potential after the electroweak symmetry breaking. We derive the
axion effective potential at one loop and we investigate the
phenomenological implications of the couplings $\sim \epsilon
\Lambda_c^2|H|^2\cos (\frac{\phi}{f_a})$ combined with the higher
order non-renormalizable operators. As we demonstrated, depending
on the value of the mass scale $\Lambda_c$, the implied changes in
the misaligned axion effective potential are drastic. For all the
cases, we took into account the constraints imposed by the LHC
experiment on the branching ratio of the Higgs to an invisible
electroweak singlet scalar hidden sector, and also the
thermalization constraints, since we need the axion to be a
non-thermal relic. The case $\Lambda_c\sim \mathcal{O}(m_H)\,$,
with $m_H$ being the Higgs mass, resulted to undesirable features
in the effective potential since the axion potential develops a
second minimum which is energetically equivalent to the Higgs
vacuum, thus we have two competing vacua in the theory. The case
$\Lambda_c\sim \mathcal{O}(m_{\nu})$, where $m_{\nu}$ is the
neutrino mass, resulted to a theory in which the axion develops a
new minimum which is energetically much less deep compared to the
Higgs minimum and the minimum of the axion potential at the origin
is separated by a barrier from the new axion minimum. Thus a
post-electroweak era first order phase transition might occur in
the axion sector, which might induce some detectable features in
the stochastic gravitational wave energy spectrum, but we did not
proceed in depth this phenomenological possibility. Finally, the
case $\Lambda_c\sim \mathcal{O}(10^{-20})\,$eV resulted to a
phenomenologically interesting situation. The axion develops a new
minimum in its effective potential which is energetically more
favorable than the minimum at the origin and with no barrier
existing between the two vacua. As the axion now is destabilized,
the axion oscillations at the origin are disturbed and the axion
is free to roll down to its potential towards the new minimum of
this potential. This can be done either in a slow-roll or
fast-roll way. Since this roll of the axion occurs during the
radiation domination era, the background EoS of the Universe might
be affected by this rolling of the axion, because the axion
composes the dark matter and this rolling affects somewhat the
total EoS making it slightly larger (fast-roll) or smaller
(slow-roll) than the radiation domination value $w=1/3$. If we
assume that this rolling of the axion occurs during the radiation
domination for frequencies probed by the Einstein Telescope, the
deformation of the background EoS might have detectable effects on
the energy spectrum of the primordial gravitational waves,
resulting to a broken power-law multiplication factor in the total
energy spectrum of the primordial gravitational waves. We examined
the energy spectrum of the primordial gravitational waves for two
inflationary scenarios, one with red-tilted tensor spectral index
and one with blue-tilted tensor spectral index for two total
background EoS parameters, $w=0.25$ and $w=0.35$. As we showed in
Fig. \ref{plot7a}, the energy spectrum of the primordial
gravitational waves remains undetectable for the red-tilted
inflationary era, without the broken power-law above the Einstein
Telescope frequencies, however for a broken power-law scenario
above the Einstein Telescope frequencies with a deformed
background EoS equal to $w=0.25$, the energy spectrum of the
primordial gravitational waves can be detectable by the Litebird
experiment as can be seen in Fig. \ref{plot6}. However, for the a
deformed background EoS equal to $w=0.35$, the energy spectrum of
the primordial gravitational waves is undetectable by all the
future experiments. Regarding the blue-tilted inflationary
theories with $n_{\mathcal{T}}=0.37$ and $r=0.003$, the signal for
a broken power-law above the Einstein Telescope frequencies, with
a deformed background EoS equal to $w=0.25$, the signal can be
detectable only by the Litebird, and the SKA for a small reheating
temperature, by Litebird, LISA, BBO and DECIGO for intermediate
reheating temperatures, and by Litebird, LISA, BBO, DECIGO and the
Einstein Telescope for a large reheating temperature, as it can be
seen in Fig. \ref{plot8}. Finally, for $w=0.35$ the signal is
undetectable for a small reheating temperature, detectable by BBO
and DECIGO for intermediate reheating temperatures, and detectable
by BBO, DECIGO and the Einstein Telescope for a large reheating
temperature, as it can be seen in Fig. \ref{plot9}. This resulting
picture offers a unique stochastic gravitational wave pattern
which can be verified by the synergy of all the future
gravitational wave experiments. As a last task, we included the
quantum corrections to the axion potential induced by the Higgs
loops and investigated the phenomenological implications of the
quantum correction terms. Phenomenologically the physical picture
is the same for all the studied cases, however a mentionable
feature is the induction of a constant vacuum energy term in the
theory. Another interesting feature of the theory we presented in
this paper is that in the case that the non-perturbative coupling
is of the order $\Lambda_c\sim \mathcal{O}(m_H)$ and if this
coupling exists primordially, due to the axion misalignment
mechanism, the symmetry breaking term of the Higgs field
$m_H^2|H|^2$ may be dynamically generated, even if this term is
primordially absent.

What we did not include in this study is the possibility of having
axion slow-roll eras during the matter domination epoch and
specifically after the matter-radiation equality. This is
possible, and if it occurs and the axion actually slow-rolls down
its potential, we may have some sort of brief early dark energy
eras during the matter domination epoch. Interestingly enough in
the literature there exist hints of an early epoch of acceleration
at a redshift $z\sim 2000$ \cite{Ye:2020btb} and these anti-de
Sitter (AdS) vacua might provide a resolution of the Hubble
tension problem. In fact, in our scenario, the axion might pass
through many distinct slow-roll AdS phases, and it is also
fascinating that this picture coincides with the phenomenology
described by the authors of Ref. \cite{Ye:2020btb} which provides
possible ways to alleviate the Hubble tension problem.
Furthermore, such slow-roll AdS eras caused on the total
background EoS during the matter domination era, might be the
source of late-time dark matter density fluctuations which can be
the seeds for large scale structure, along with the inflationary
modes \cite{Frieman:1995pm}. These features are quite interesting
from a phenomenological point of view and should be further
investigated in a focused future work.

\section*{Acknowledgments}

This research has been is funded by the Committee of Science of
the Ministry of Education and Science of the Republic of
Kazakhstan (Grant No. AP19674478).

\end{document}